\documentclass[sigconf]{acmart}
\usepackage{graphicx}
\usepackage{booktabs}
\usepackage{hyperref}
\usepackage{algorithm}
\usepackage{algpseudocode}
\usepackage{multirow}
\usepackage{float}
\usepackage{footnote}
\usepackage{color}
\usepackage{graphicx} 
\usepackage{microtype}
\usepackage{graphicx}
\usepackage{booktabs}  
\usepackage[table]{xcolor}  
\usepackage{amsfonts}

\usepackage{caption}
\usepackage{subcaption}
\usepackage{tikz}
\usepackage{multirow}
\usepackage{array}
\usepackage{enumitem}
\usepackage{mdframed}
\usepackage{tcolorbox}

\begin{document}

\setcopyright{none}
\settopmatter{printacmref=false} 
\renewcommand\footnotetextcopyrightpermission[1]{} 

\thispagestyle{empty}  

\title{Cluster-based Adaptive Retrieval: Dynamic Context Selection for RAG Applications}

\author{Yifan Xu}
\affiliation{
\institution{Coinbase}
\city{Mountain View}
  \state{CA}
  \country{USA}
}

\author{Vipul Gupta}
\authornote{Work done while at Coinbase.}
\affiliation{
\institution{Coinbase}
\city{Bengaluru}
  \state{KA}
  \country{India}
}

\author{Rohit Aggarwal}
\affiliation{
\institution{Coinbase}
\city{Bengaluru}
  \state{KA}
  \country{India}
}

\author{Varsha Mahadevan}
\affiliation{
\institution{Coinbase}
\city{Bengaluru}
  \state{KA}
  \country{India}
}

\author{Bhaskar Krishnamachari}
\authornote{Bhaskar Krishnamachari is a paid consultant for Coinbase
 and has assisted on this paper in this capacity.}
\affiliation{
\institution{University of Southern California}
\city{Los Angeles}
  \state{CA}
  \country{USA}
}

\renewcommand{\shortauthors}{Xu, et al.}


\begin{abstract}
Retrieval-Augmented Generation (RAG) enhances large language models (LLMs) by pulling in external material, document, code, manuals, from vast and ever-growing corpora, to effectively answer user queries. The effectiveness of RAG depends significantly on aligning the number of retrieved documents with query characteristics: narrowly focused queries typically require fewer, highly relevant documents, whereas broader or ambiguous queries benefit from retrieving more extensive supporting information. However, the common static top-k retrieval approach fails to adapt to this variability, resulting in either insufficient context from too few documents or redundant information from too many.

Motivated by these challenges, we introduce Cluster-based Adaptive Retrieval (CAR), an algorithm that dynamically determines the optimal number of documents by analyzing the clustering patterns of ordered query-document similarity distances. CAR detects the transition point within similarity distances, where tightly clustered, highly relevant documents shift toward less pertinent candidates, establishing an adaptive cut-off that scales with query complexity. On Coinbase’s CDP corpus and the public MultiHop-RAG benchmark, CAR consistently picks the optimal retrieval depth and achieves the highest TES score, outperforming every fixed top-k baseline. In downstream RAG evaluations, CAR cuts LLM token usage by 60\%, trims end-to-end latency by 22\%, and reduces hallucinations by 10\% while fully preserving answer relevance. Since integrating CAR into Coinbase’s virtual assistant, we’ve seen user engagement jump by 200\%.
\end{abstract}


\maketitle

\section{Introduction}

\begin{figure}[t] 
    \centering
    \includegraphics[width=0.47\textwidth]{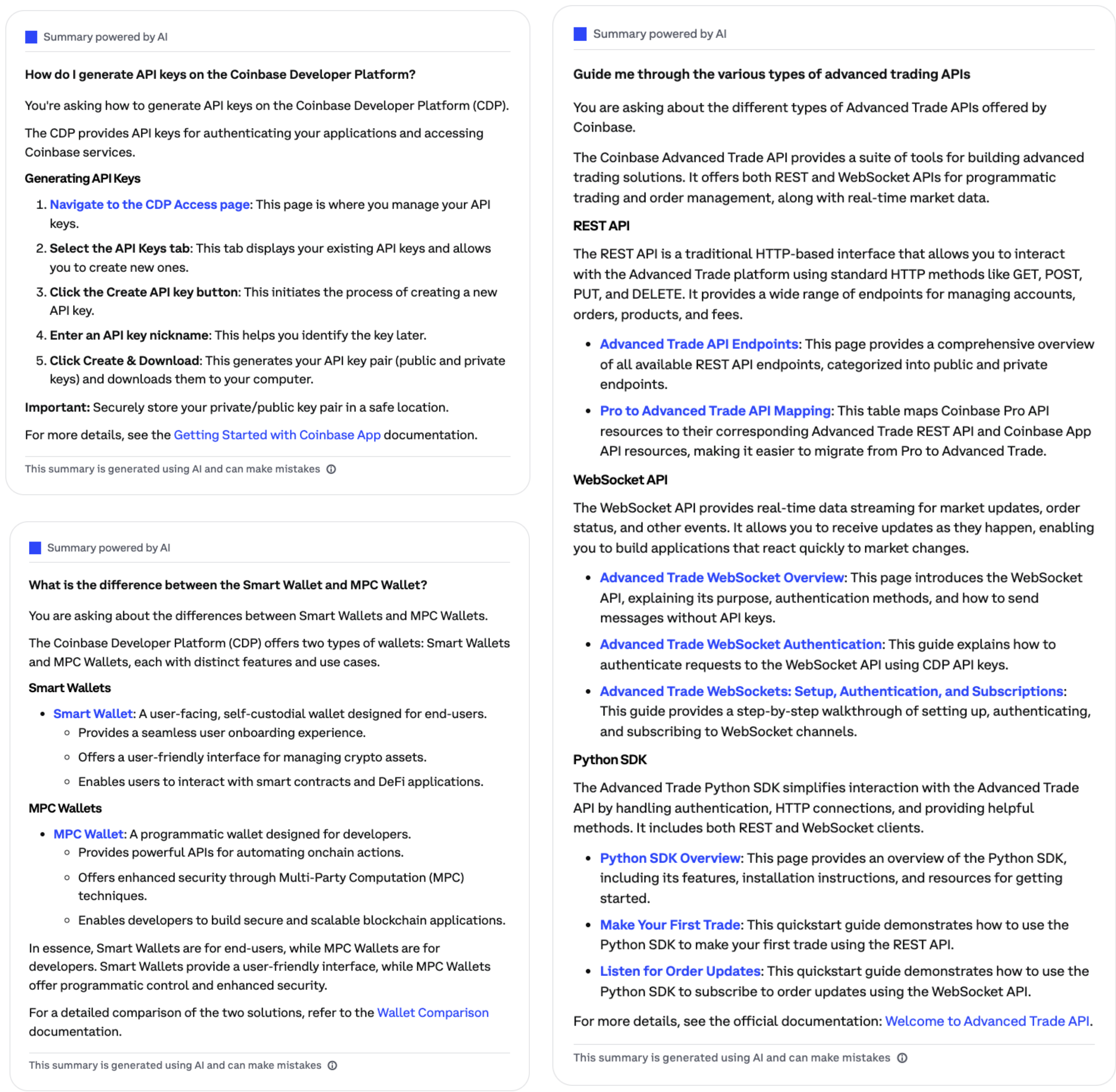}
    \caption{
    Examples illustrating varying query complexity in CDP documentation search: (top-left) simple query answered by one API doc; (bottom-left) moderate query needing multiple docs to clarify product relationships; (right) complex query requiring synthesis across diverse sources.
    }
    \label{fig:front_end} 
    \Description{CDP Example Queries}
\end{figure}

Recent advances in large language models (LLMs) \cite{achiam2023gpt,team2024gemini} have catalyzed a paradigm shift in natural language processing, with Retrieval-Augmented Generation (RAG) \cite{lewis2020retrieval, jiang2023active, chen2024benchmarking} emerging as a prominent framework for question answering systems—powering applications such as retrieval-based answer generation and customer service chatbots. RAG systems enhance language model performance by integrating external knowledge — drawn from continuously evolving and heterogeneous sources — directly into the generation process. These RAG systems typically relies on a static top-\(k\) retrieval strategy \cite{lewis2020retrieval,gao2023retrieval}: the query is first encoded into an embedding vector, and then the \(k\) documents whose embeddings are most similar to the query are selected as context. Although effective for straightforward information needs, this approach assumes that every query benefits equally from an identical number of documents in a production system—a presumption increasingly at odds with the varied and dynamic nature of real-world content.

\begin{figure}[t] 
    \centering
    \includegraphics[width=0.47\textwidth]{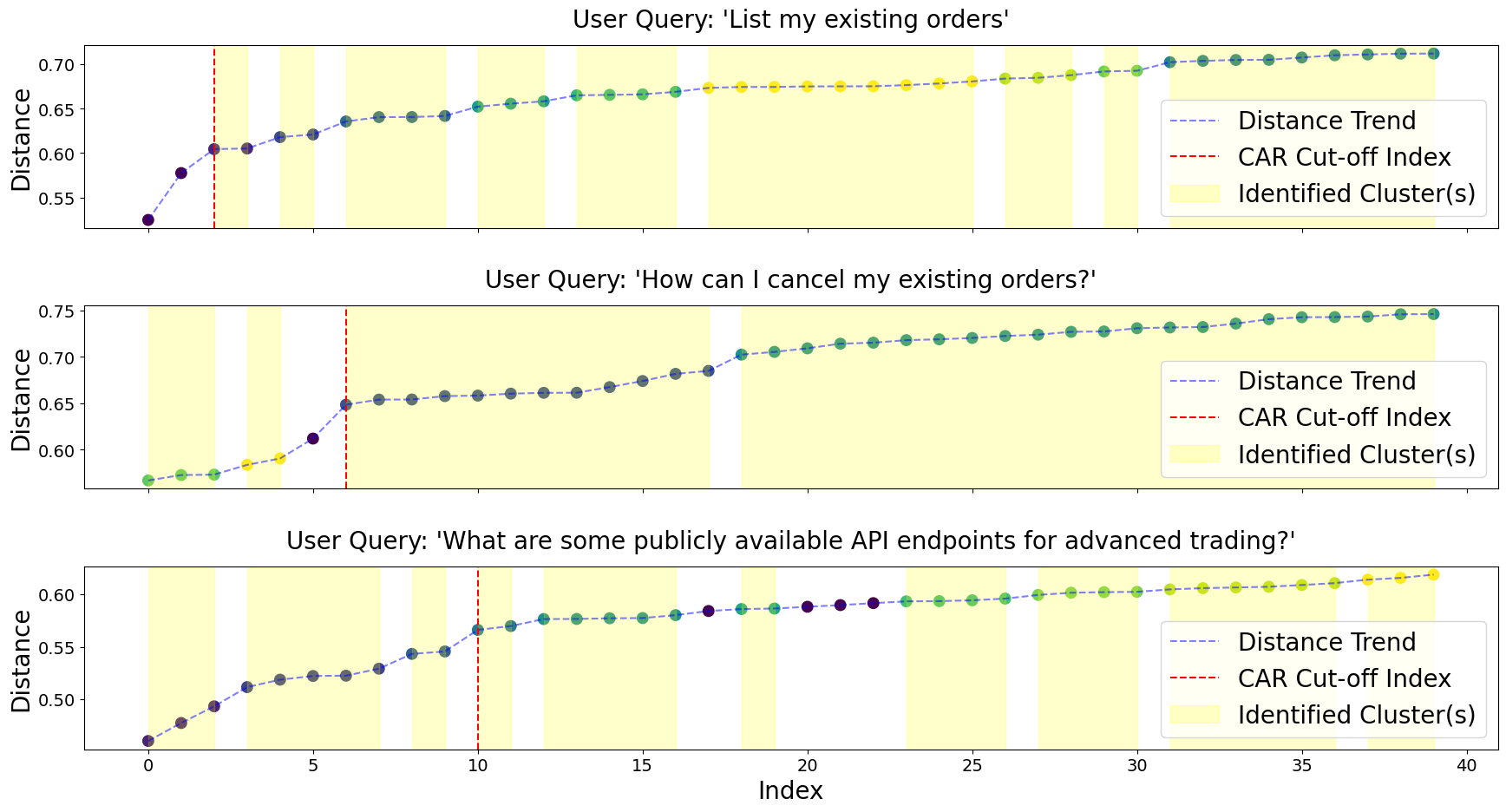}
    \caption{Visualization of CAR’s cutoff mechanism: it adaptively selects retrieval thresholds (e.g., 3, 7, 11) based on clustering patterns in the embedding space, preserving relevant documents and filtering out weaker ones. The clustering outcome—shaped by query complexity and document characteristics—guides the cutoff decision.}
    \Description{CAR cutoff mechanism}
    \label{fig:cutoff} 
\end{figure}
This challenge is particularly evident on our Coinbase developer platform\footnote{\url{https://coinbase.com/developer-platform}}. As shown in Figure~\ref{fig:front_end}, user queries range from simple fact-based lookups to complex multi-document understanding tasks. For instance, a developer seeking to ``\textit{generate an API key}" may require only a single, well-defined snippet from the API documentation. In contrast, a more intricate query—such as understanding ``\textit{the differences between Smart Wallets and MPC Wallets}"—necessitates retrieving documents from both products to provide a comprehensive response. Even more complex queries, such as ``\textit{Guide me through the various types of advanced trading APIs}", require retrieving and synthesizing information from multiple diverse sources. Answering such a query involves referencing the REST API documentation for available endpoints, the WebSocket API guide for real-time updates and authentication methods, and the Python SDK documentation for programmatic interaction with the trading APIs. These examples highlight the inherent trade-offs in retrieval: while simpler queries can be effectively addressed with a minimal document set, more complex inquiries demand an adaptive retrieval approach to balance completeness against computational efficiency.

At Coinbase, our RAG-based pipeline forms the backbone of our virtual assistant systems integrated into the documentation search bar \footnote{\url{https://docs.cdp.coinbase.com/}}and the active chatbot on our CDP Discord Server \footnote{\url{https://discord.com/invite/cdp}}. These systems support developers as they navigate our comprehensive and diverse documentation, which includes API references\footnote{API Example: \url{https://docs.cdp.coinbase.com/advanced-trade/reference}}, tabular data\footnote{Tabular Example: \url{https://docs.cdp.coinbase.com/advanced-trade/docs/rest-api-pro-mapping}}, markdown texts with embedded code\footnote{Code Example: \url{https://docs.cdp.coinbase.com/advanced-trade/docs/rest-api-auth}}, and GitHub source code\footnote{Github Example: \url{https://github.com/coinbase/coinbase-advanced-py}}. We observe that a fixed retrieval strategy often falls short in balancing the need for detailed, diverse information against efficiency concerns. This insight has motivated us to pursue a more adaptive retrieval method that dynamically tailors the number of documents based on the complexity of each query, without incurring a latency penalty in production environments.

To address these challenges, we introduce Cluster-based Adaptive Retrieval (CAR), a method that dynamically adjusts the number of documents retrieved by combining the natural ordering of similarity distances with a clustering analysis. CAR ranks candidate documents by their similarity distance, where lower distances imply higher relevance, and then applies a clustering algorithm to these ordered values. This approach groups documents with similar relevance patterns (each cluster potentially capturing a different facet of the query) and reveals natural breakpoints in the distance distribution. For example, A dense cluster of low-distance documents followed by a sharp increase indicates that only a few documents are truly pertinent, while a gradual transition across clusters suggests that a broader set is needed. Using these insights, CAR adaptively determines a cut-off point (illustrated in Figure \ref{fig:cutoff}), ensuring that the language model receives an optimally sized set of relevant documents, enhancing the quality of the response without incurring unnecessary retrieval overhead.

\begin{figure}[t] 
    \centering
    \includegraphics[width=0.43\textwidth]{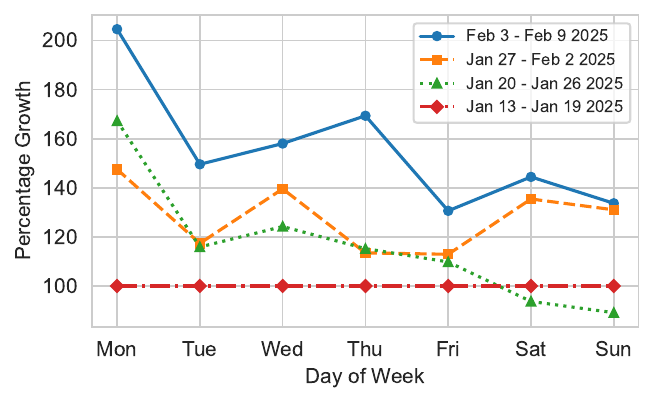}
    \caption{Week-over-week growth in CDP search queries after deploying the AI summary system, showing increasing user engagement (week of January 13-19 is taken as reference).}
    \label{fig:amplitude} 
    \Description{CDP Search week on week growth}
\end{figure}

On December 16, 2024, the CDP documentation search bar transitioned from keyword-based search to a RAG pipeline utilizing our CAR algorithm. Figure~\ref{fig:amplitude} shows weekly query growth from January 13 to February 9, 2025, using the initial week (January 13–19) as the baseline to ensure query volume stabilization. Query volume exhibited steady growth, peaking above 200\% on Monday during the week of February 3–9, 2025, and maintaining robust engagement throughout the week. This pattern highlights increased user adoption and perceived value of the AI-driven retrieval system.\footnote{Absolute query counts are omitted for confidentiality; only relative growth trends are presented.}

Our extensive evaluations of CAR span two distinct scenarios: (1)~real-world deployment on Coinbase's large-scale, unstructured, and continuously updated corpora, and (2)~the MultiHop-RAG news dataset~\cite{tang2024multihop}, chosen specifically for its domain-agnostic content and diverse retrieval complexity (queries requiring between one and four documents). Experimental results consistently demonstrate CAR's superior performance over the widely-used fixed top-\(k\) cut-off approach. In particular, CAR retrieves more precisely targeted document sets for narrowly scoped queries while adaptively broadening context for complex inquiries—all without additional inference steps or modeling overhead. Furthermore, CAR's computational cost is negligible, introducing less than 50 milliseconds of latency per query. Consequently, CAR effectively addresses retrieval challenges identified in our CDP documentation scenario, establishing itself as a practical, scalable, and robust solution for advanced RAG-based systems.

\vspace{-2mm}
\section{Related Work}
\label{sec:related_works}

Next, we discuss several related approaches that focus on retrieval quality and propose CAR as a novel and efficient alternative.

\textbf{Retrieval-Augmented Generation (RAG)}:  The seminal work by \cite{lewis2020retrieval} introduced RAG as a means to augment large language models (LLMs) with external knowledge. In traditional RAG systems, a fixed top-\textit{k} strategy is used to retrieve documents from an external corpus. Although effective in some scenarios, this approach can lead to 
either redundancy or omission of crucial information when dealing with diverse and complex queries.

Recent research has sought to address these limitations through dynamic retrieval strategies. For example, DRAGIN \cite{su2024dragin} and MemoRAG \cite{memorag2024} adaptively select documents based on the model’s real-time information needs or memory-inspired signals, while LongRAG \cite{longrag2024} and ChatQA 2 \cite{chatqa22023} focus on handling long-context scenarios. Despite these advances, many of these methods still rely on multi-stage processing or are tailored to specific applications, motivating the need for a more general and lightweight dynamic retrieval mechanism.

\textbf{Dynamic Cutoff Algorithms}:
Dynamic cutoff algorithms aim to determine the optimal number of documents to retrieve on a per-query basis. Several recent approaches, such as Self-RAG~\cite{selfrag2023} and SeaKR~\cite{seakr2023}, leverage self-reflection or uncertainty estimates from LLMs to trigger retrieval when needed. Similarly, RA-DIT \cite{radit2023} integrates dual instruction tuning to refine retrieval output, and RETRO \cite{borgeaud2022improving} dynamically adjusts context windows to improve efficiency on massive corpora. Although these methods offer promising directions, they often introduce extra computational overhead in terms of added finetuning, training or inference stages, increase the overall complexity of the system. 

\textbf{Re-ranking Methods}:
Re-ranking techniques have been widely employed to improve the quality of retrieved documents. For example, DSLR \cite{hwang2024dslr} refines retrieval results by decomposing documents into sentences, filtering irrelevant content, and reconstructing coherent passages. RankRAG \cite{rankrag2023} and Re2G \cite{glass2022re2g} utilize advanced ranking models to reorder candidates according to relevance. Although these methods improve contextual quality, they often use a rerank model incur additional computational steps that can increase latency, cost and complexity of the system.


Unlike the approaches described above, CAR dynamically determines the optimal number of documents by identifying natural breakpoints in the similarity score distribution via clustering. Rather than relying on fixed top-\textit{k} selection \cite{lewis2020retrieval} or additional uncertainty-based retrieval triggers \cite{selfrag2023, seakr2023}, CAR automatically adapts to the complexity of each query. By discarding clusters of less relevant documents, CAR ensures that only the most pertinent context is passed to the language model. This adaptive mechanism reduces noise and hallucination while avoiding the overhead of extra re-ranking or dual-stage tuning methods (e.g., RA-DIT \cite{radit2023}). As such, CAR enables efficient and adaptive retrieval even in complex, heterogeneous document repositories, making it particularly well-suited for real-time applications.

\begin{figure*}[t] 
    \centering
    \includegraphics[width=0.9\textwidth]{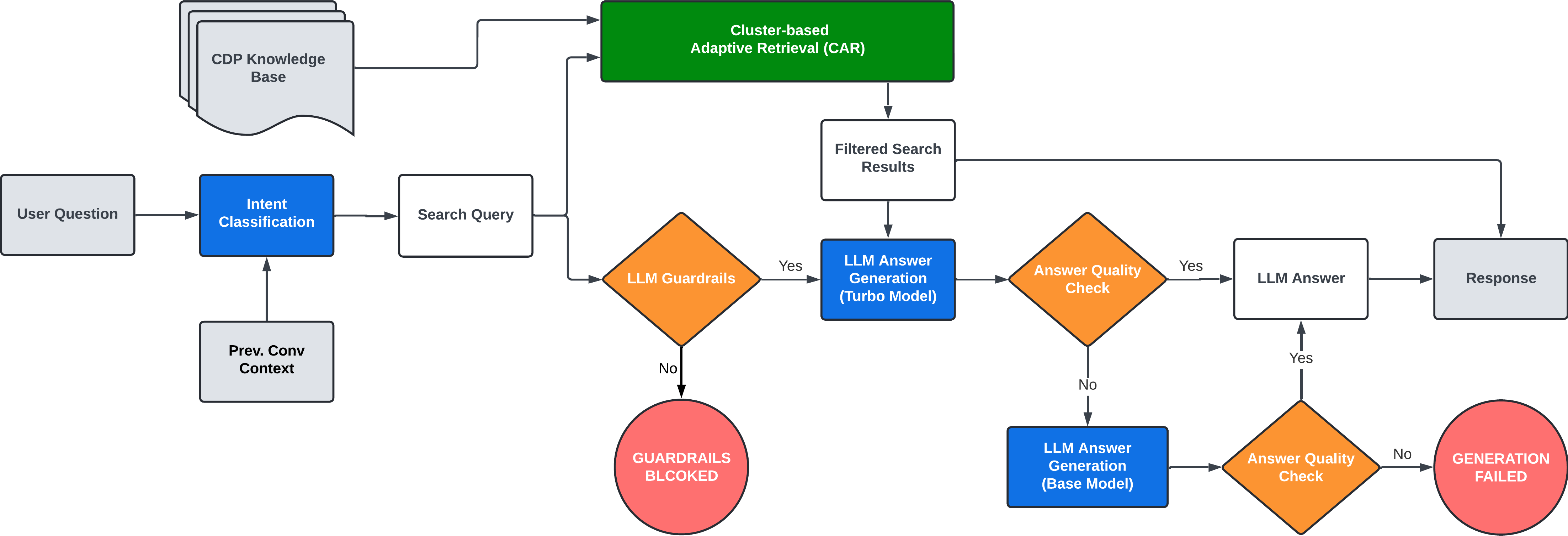} 
    \caption{Overview of the CDP RAG system pipeline, from query input to response generation. It employs multiple LLM calls for intent detection, retrieval, guardrails, answer generation, and quality checks, ensuring accuracy and trust. The CAR algorithm enhances search result relevance.}
    \Description{CDP Pipeline}
    \label{fig:pipeline} 
\end{figure*}


Next, we describe the deployed application and cluster-based approach of CAR in more detail. 

\section{Deployed RAG Application: CDP Docs Search and Discord Chatbot}
We developed a RAG framework that powers both a documentation search interface and a customer service chatbot for the CDP documentation. This compound system leverages multiple LLM calls—including intent detection, dedicated input guardrails, and iterative quality checks—to ensure that users receive fast, reliable, and high-quality answers. Our search interface meets strict latency requirements, with an average response time of 5 seconds (standard deviation: 1 second). In addition to search functionality, the chatbot is tailored for advanced developer assistance, supporting multi-turn interactions, contextual refinement, and code snippet debugging. At a high level, the architecture, as illustrated in Figure \ref{fig:pipeline}, processes user queries through a structured pipeline that retrieves relevant documents and generates responses using large language models (LLMs). The key components of the pipeline include:

\textbf{1. Query Processing and Dynamic Document Retrieval:}
When a user submits a question, the system converts the input into query embeddings and performs a semantic search against a vector database of pre-embedded CDP documents. The proposed Cluster-based Adaptive Retrieval (CAR) algorithm then dynamically filters the initial set of candidate documents, ensuring that only the most pertinent content is forwarded to subsequent stages.

\textbf{2. Answer Generation:}
A fast, low-latency language model, referred to as the Turbo model, is used first to generate an initial response. This draft is subjected to a series of output quality checks (outlined below). If the response fails to meet the required standards, the system triggers a fallback mechanism: the same query and context are sent to a more capable, higher-latency base model for re-generation. This ensures that every response is swift and adheres to rigorous standards of accuracy, safety, and grounding.

\textbf{3. Input Guardrails and Output Quality Checks:}
Before answer generation, the system enforces input guardrails on both user query and retrieved documents to ensure safety and suitability of the content. After response generation, output quality checks are applied to verify that the answer meets the following criteria:
\begin{itemize}[leftmargin=13pt]
\item \textit{Safety:} The answer must be free of harmful or inappropriate content. 
\item \textit{Trustworthiness:} The response should reliably align with verified information. \item \textit{Groundedness:} The answer must be well-supported and properly anchored in the retrieved documents. 
\end{itemize}

\textbf{4. Multi-turn Chatbot Enhancements:}
For conversational queries, the chatbot extends the document search pipeline with multi-turn context processing. Each query is evaluated alongside the preceding conversation history (while discarding previously retrieved documents to maintain clarity). An intent classifier analyzes the multi-turn dialogue to refine document retrieval and tailor the LLM response accordingly.

By tailoring retrieval, enforcing robust safeguards, and incorporating multi-turn context, our solution meets evolving developer needs while upholding strict latency and quality standards.

\section{Method}
\label{sec:method}

We introduce \textit{Cluster-based Adaptive Retrieval (CAR)}, a technique that replaces static top-\(k\) document selection with a data-driven approach based on clustering patterns observed in the ranked similarity distances between the query and candidate documents. Below, we describe the key steps of CAR, highlight its motivation in the context of Retrieval-Augmented Generation (RAG), and explain why this methodology is particularly suited for complex, evolving knowledge sources of mixed content (e.g., code, documentation, FAQs).

\subsection{Overview of the Algorithm}
\begin{algorithm}
\caption{Cluster‑based Adaptive Retrieval (CAR)}
\label{alg:car_algorithm}
\begin{algorithmic}[1]
\State \textbf{Input:} Query $q$, corpus $\mathcal{D}$,
      hyper‑parameter grid $\Theta$   \Comment{$\Theta$ can include $k$, \texttt{eps}, \texttt{linkage}, …}
\State \textbf{Output:} Optimised subset of retrieved documents $\mathcal{R}$, cutoff point $C$, best hyper‑parameters $\theta^\ast$
\vspace{0.2em}

\Procedure{CAR}{$q, \mathcal{D}, \Theta$}
    \State $\mathcal{R} \gets \Call{RetrieveTopN}{q, \mathcal{D}, N}$ \Comment{Retrieve top‑$N$ candidates}
    \State $\{(d_1,\delta_1),\dots,(d_N,\delta_N)\} \gets \mathcal{R}$  \Comment{Distances $\delta_n$ to the query}

    \For{$n=1$ \textbf{to} $N$}
        \State $\tilde{\delta}_n \gets
          \dfrac{\delta_n-\min(\delta_1,\dots,\delta_N)}
                {\max(\delta_1,\dots,\delta_N)-\min(\delta_1,\dots,\delta_N)}$
    \EndFor

    \State $s^\ast \gets -\infty$, $\theta^\ast \gets \text{None}$
    \ForAll{$\theta \in \Theta$}        \Comment{Grid search over candidate settings}
        \State $\mathcal{C}_\theta \gets 
               \Call{Cluster}{\{(\tilde{\delta}_n,n)\}_{n=1}^{N}, \theta}$
        \State $s_\theta \gets \Call{SilhouetteScore}{\{(\tilde{\delta}_n,n)\}, \mathcal{C}_\theta}$
        \If{$s_\theta > s^\ast$}
            \State $s^\ast \gets s_\theta$, $\theta^\ast \gets \theta$
        \EndIf
    \EndFor

    \State $\mathcal{C} \gets
           \Call{Cluster}{\{(\tilde{\delta}_n,n)\}_{n=1}^{N}, \theta^\ast}$
    \State Identify contiguous cluster regions $\mathcal{P} \subset \mathcal{C}$ \Comment{Detect stable clusters along ranking}

    \State $S \gets \Bigl\{\, i \in \{2,\dots,N\}\;:\;
                \mathcal{C}_i \neq \mathcal{C}_{\,i-1} \Bigr\}$
          \Comment{$\mathcal{C}_i$ = cluster label of the $i$‑th document; Each $i\!\in\!S$ is the \emph{first} rank of every new cluster region}
    
    \If{$S = \varnothing$}                      \Comment{All $N$ docs fell into one region}
        \State $C \gets N$                      \Comment{→ no natural cutoff; keep everything}
    \Else
        \State $G \gets \bigl\{\tilde{\delta}_i - \tilde{\delta}_{\,i-1}\;|\;i \in S\bigr\}$
               \Comment{Gradients at boundaries}
        \ForAll{$i \in S$}
            \State $score_i \gets
                   \dfrac{g_i}{\max_{j \in S} g_j} \;+\;
                   \dfrac{i}{N}$                \Comment{Gap size + rank penalty}
        \EndFor
        \State $i^\ast \gets \arg\max_{i \in S} score_i$
        \State $C \gets i^\ast - 1$ \Comment{Cutoff = last doc before best gap}
    \EndIf
    \State $\mathcal{R} \gets \{\, d_i \mid i \leq C \,\}$
    \State \Return $\mathcal{R}, C, \theta^\ast$
\EndProcedure
\end{algorithmic}
\end{algorithm}

Let $\mathcal{D}$ denote a large-scale corpus of documents, and let $q$ be a query encoded into the same embedding space as the documents in $\mathcal{D}$. In Algorithm \ref{alg:car_algorithm}, we present CAR method proceeds in three main phases:

\textbf{Phase 1: Initial Retrieval.}
We begin by performing a standard embedding-based lookup against $\mathcal{D}$ to retrieve the top-$N$ candidate documents most similar to $q$. This step yields a ranked list
\begin{equation}
\label{eq:candidate_set}
\{(d_1, \delta_1), (d_2, \delta_2), \ldots, (d_N, \delta_N)\},
\end{equation}
where each $d_n$ is a candidate document and $\delta_n$ is its distance (or inverse similarity) from $q$, with $\delta_1 \leq \delta_2 \leq \cdots \leq \delta_N$. The parameter $N$ is chosen as an upper bound large enough to capture a diverse set of possibly relevant results. In practice, we set n where it can include all correct documents for at least 90+\% queries. Here, smaller values of $\delta_n$ indicate higher similarity.

Before proceeding with clustering, we normalize these distance scores to ensure consistency in scale. Specifically, we define the normalized distance $\tilde{\delta}_n$ as follows:
\[
\tilde{\delta}_n = \frac{\delta_n - \min\{\delta_1, \dots, \delta_N\}}{\max\{\delta_1, \dots, \delta_N\} - \min\{\delta_1, \dots, \delta_N\}},
\]
which rescales each $\delta_n$ to lie in the range $[0,1]$. This normalization facilitates the detection of natural gaps in the ranked list.

\textbf{Phase 2: Cluster Formation.}
From the ranked list, CAR applies a clustering algorithm—e.g., HDBSCAN \cite{mcinnes2017hdbscan}, OPTICS~\cite{ankerst1999optics}, or BisectingKMeans~\cite{karypis2000comparison}—to partition the $\delta_N$ candidates into cohesive segments with small intra‑cluster distances. Any method that (i) tolerates varied cluster shapes and (ii) supports explicit outlier labelling can be used, as these properties naturally capture the ``plateaus" in relevance.

When the chosen algorithm has hyper‑parameters that influence the final partition-for example, the number of clusters $k$ in BisectingKMeans or the neighbourhood radius $\epsilon$ in DBSCAN—we determine the optimal setting by maximising the average \textit{silhouette score}~\cite{rousseeuw1987silhouettes,shahapure2020cluster}. Concretely, CAR performs a grid search over the user‑specified hyper‑parameter space and retains the configuration that yields the highest silhouette value; full details of this procedure are given in Section~\ref{hyperparameter-search}.

\textbf{Phase 3: Adaptive Cutoff Selection.}
Following the identification of clusters, CAR sets a dynamic cutoff $C \leq K$ to finalize which candidates should be included. This step capitalizes on the fact that dense clusters often sit near the top of the ranked list, where documents are most relevant to $q$. Concretely, CAR determines $C$ based on cluster boundaries, discarding documents that fall outside the highly relevant clusters. By aligning the cutoff with inherent gaps in the distance distribution, CAR adaptively tailors the retrieval set size to the complexity of the query.

\textit{Cutoff Score with Position Penalty:}  
In our implementation, we apply an additional position penalty to adjust the gap score based on its rank, thereby discouraging an overly early cutoff. Let \(S\) denote the set of indices where a new cluster begins (i.e., the boundaries between clusters). For each \(i \in S\), define the gap at the boundary as
\[
g_i = \tilde{\delta}_i - \tilde{\delta}_{i-1}.
\]
We then define the composite score for a gap at a boundary index \(i\) as:
\[
\text{score}_i = \frac{g_i}{\displaystyle \max_{j \in S} g_j} 
+ \frac{i}{N} 
\quad 
\forall~i \in S,
\]
where the second term (\(\tfrac{i}{N}\)) represents a position-based ``penalty'' that discourages cutoffs too close to the top of the list.

The gap term \(\tfrac{g_i}{\max_{j \in S} g_j}\) captures the relative magnitude of the boundary, highlighting where a large discontinuity separates adjacent clusters. Simultaneously, \(\tfrac{i}{N}\) accounts for the position of that boundary in the overall ranking. This is particularly helpful for \emph{long-tail} or more open-ended queries where multiple reasonably large gaps may appear. By adding \(\tfrac{i}{K}\), the method adaptively shifts the cutoff further down the ranked list if multiple boundary gaps are comparable in size, ensuring we include enough relevant documents. 

Once each boundary index \(i\) has a composite cutoff \(\text{score}_i\), we select the index \(i^*\) that maximizes this score:
\[
i^* = \arg\max_{i \in S} \text{score}_i,
\]
and set the cutoff as
\[
C = i^* - 1.
\]
Hence, the final retrieval set includes all documents up to the candidate preceding the highest-scoring boundary. This approach balances gap size and ranking position, yielding a flexible mechanism that tailors itself to the intrinsic structure of the candidate set rather than a one-size-fits-all threshold.

\begin{table}[t]
    \centering
    \resizebox{0.4\textwidth}{!}{%
    \begin{tabular}{lccc}
        \toprule
        \textbf{Method} & \textbf{Accuracy ↑} & \textbf{Avg. Candidates ↓} & \cellcolor{gray!20} \textbf{TES ↑} \\
        \midrule
        \multicolumn{4}{c}{\textbf{Clean Query 300}} \\
        \midrule
        Top-3  & 0.97  & 3.0  & \cellcolor{gray!20} 0.700 \\
        Top-5  & 0.99  & 5.0  & \cellcolor{gray!20} 0.553 \\
        Top-10 & 1.00  & 10.0 & \cellcolor{gray!20} 0.417 \\
        \hline
        CAR (Ours) & 0.98  & 2.1  & \cellcolor{gray!20} \textbf{0.866} \\
        \midrule
        \multicolumn{4}{c}{\textbf{Noisy Query 300}} \\
        \midrule
        Top-3  & 0.60  & 3.0  & \cellcolor{gray!20} 0.433 \\
        Top-5  & 0.67  & 5.0  & \cellcolor{gray!20} 0.374 \\
        Top-10 & 0.87  & 10.0 & \cellcolor{gray!20} 0.363 \\
        \hline
        CAR (Ours) & 0.69  & 3.5  & \cellcolor{gray!20} \textbf{0.459} \\
        \bottomrule
    \end{tabular}%
    }
    \caption{Comparison of retrieval accuracy and efficiency for different retrieval methods across Clean and Noisy Queries. TES highlights the trade-off between accuracy (higher is better) and the number of candidates returned (lower is better).}
    \label{tab:search_experiment_cdp}
\end{table}

\begin{table*}[htbp]
\centering
\resizebox{0.9\textwidth}{!}{
\begin{tabular}{|l|c|c|c|c|}
\hline
\textbf{Method} & \textbf{BAAI/llm-embedder}~\cite{llm_embedder} & \textbf{text-embedding-ada-002}~\cite{openai2023embedding} & \textbf{text-embedding-3-small}~\cite{openai2024embedding} & \textbf{text-embedding-3-large}~\cite{openai2024embedding} \\
\hline
Top-3 & 0.17 & 0.25 & 0.26 & 0.25 \\
Top-5 & 0.22 & 0.30 & 0.31 & 0.29 \\
Top-10 & 0.25 & 0.33 & 0.32 & 0.30 \\
Top-20 & 0.25 & 0.29 & 0.30 & 0.29 \\
\hline
CAR (Ours) & \textbf{0.25} & \textbf{0.34} & \textbf{0.34} & \textbf{0.31} \\
\hline
\end{tabular}
}
\caption{Trade-off Efficiency Score (TES; higher is better) on the MultiHop-RAG dataset.  
CAR consistently surpasses fixed top-\(k\) retrieval across four leading embedding models while using fewer documents on average.}
\label{tab:search_experiment_multihop}
\end{table*}

\section{Experiments}
This section answers three practical questions.  
First, \emph{does} the proposed \textit{Cluster-based Adaptive Retrieval} (CAR) algorithm achieve a better balance between accuracy and number of candidates retrieved when each query has only one gold document?
Second, \emph{is} that improvement robust to different corpus domain and embedding choice when the answer requires reasoning over multiple pieces of evidence?  
Third, \emph{what} is the end-to-end impact on a production Retrieval-Augmented Generation (RAG) pipeline—namely, response relevancy, hallucination rate, token budget, and wall-clock latency?

To that end we design a two-stage protocol.

\begin{enumerate}[leftmargin=10pt]
\item\textbf{Retrieval-only stage.}  
    We benchmark CAR against fixed top-\(k\) cut-offs on two datasets of complementary structure.  
    On the Coinbase CDP corpus each query admits \textit{one} authoritative answer document; on the public MultiHop-RAG corpus~\cite{tang2024multihop} each question is resolved by \textit{2–4} documents dispersed across the collection. Accuracy and the average number of candidates retrieved are recorded, and their balance is summarized with the Trade-off Efficiency Score (TES) introduced below.
\item\textbf{Full RAG stage.}  
    We plug each retrieval strategy into Coinbase’s production search-bar assistant and replay 1400 real user queries. Besides retrieval statistics, we measure end-to-end latency and apply the \textit{DeepEval} open-source library\footnote{GitHub URL: \url{https://github.com/confident-ai/deepeval}} to score answer relevancy, contextual relevancy, and hallucination rate.
\end{enumerate}

\subsection{Retrieval Effectiveness Evaluation}

\textbf{Benchmarks.}
We evaluate on two corpora that pose complementary challenges.  
(1)~Coinbase CDP. From the \textit{Advanced Trading} REST API\footnote{\url{https://docs.cdp.coinbase.com/advanced-trade/reference}} we derive 600 queries that each map unambiguously to a single endpoint.  
To stress-test robustness, we construct two splits: a high-precision \emph{clean} set filtered by an offline embedder’s top-5 recall, and a paraphrased \emph{noisy} split whose ground-truth document is harder to recover.  
(2)~MultiHop-RAG. The news dataset of Tang \textit{et al.}~\cite{tang2024multihop} contains 2556 questions whose answers require evidence scattered across two to four documents, mirroring multi-document reasoning in production RAG systems.

\textbf{Embedding models.}
CDP experiments use \textsc{TextEmbedding Gecko}~\cite{lee2024gecko}.  
For MultiHop-RAG we probe four encoders—\texttt{BAAI llm embedder}~\cite{llm_embedder}, \texttt{text embedding ada 002}~\cite{openai2023embedding}, \texttt{text embedding 3 small}, and \texttt{text embedding 3 large}~\cite{openai2024embedding}.  

\textbf{Metrics.}
We report \emph{Accuracy} (ground-truth document present) and the \emph{Average Candidates Returned}.  
To penalise over-retrieval we introduce the \emph{Trade-off Efficiency Score} (TES):
\[
    \mathrm{TES} \;=\;
    \frac{\text{Accuracy}}
         {\ln\!\bigl(1 + \text{Average Candidates Returned}\bigr)} .
\]

\textbf{Results.}
Tables~\ref{tab:search_experiment_cdp} and \ref{tab:search_experiment_multihop} summarize findings.  
On the CDP benchmark, CAR matches or exceeds the accuracy of static top-\(k\) while returning \(\approx\!30\%\) fewer documents, yielding the highest TES.  
On MultiHop-RAG, CAR outperforms top-\(k\) across all four encoders, confirming that its adaptive cut-off generalizes beyond any single embedding space.  
Overall, CAR delivers a superior accuracy–latency trade-off that is critical for large-scale Retrieval-Augmented Generation.

\begin{table*}[t]
    \centering
    \resizebox{0.9\textwidth}{!}{%
    \begin{tabular}{lcccccc}
        \toprule
        \textbf{Method} & \textbf{\# Documents ↓} & \textbf{Tokens ↓} & \textbf{Latency (s) ↓} & \textbf{Answer Relevancy ↑} & \textbf{Contextual Relevancy ↑} & \textbf{Hallucination ↓} \\
        \midrule
        Top-16 & 16.00 & 14,896 & \textbf{5.14} & 0.87 & 0.48 & 0.16 \\
        Top-40 & 40.00 & 34,472 & 6.99 & \textbf{0.90} & 0.44 & 0.12 \\
        \hline
        CAR (Ours) & \textbf{15.57} & \textbf{13,477} & 5.46 & 0.89 & \textbf{0.49} & \textbf{0.10} \\
        \bottomrule
    \end{tabular}%
    }
    \caption{Comparison of retrieval efficiency, latency, and LLM response evaluation for Top-40, Top-16, and CAR methods averaged over 1400 user queries on the CDP search bar. Higher is better (↑) for Answer Relevancy and Contextual Relevancy, while lower is better (↓) for Docs/Tokens Consumption, Latency, and Hallucination.}
    \label{tab:retrieval_efficiency}
\end{table*}

\subsection{Full RAG Evaluation (Coinbase Deployed Application)}

The RAG evaluation is designed to reflect real-world usage where more condense relevant documents may enhance LLM generated response. A query can benefit from several relevant documents that together provide a richer and multifaceted context.

\begin{figure}[t] 
    \centering
    \includegraphics[width=0.47\textwidth]{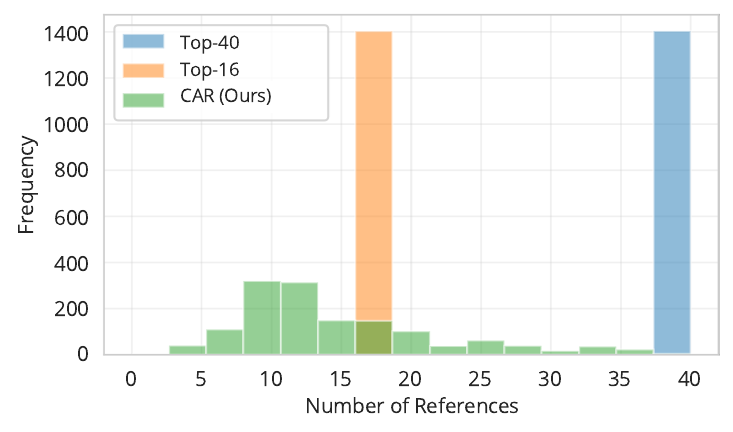}
    \caption{CAR dynamically adjusts the number of retrieved references based on query complexity, unlike fixed retrieval methods.}
    \label{fig:reference_distribution} 
    \Description{Reference Count Histogram}
\end{figure}

\textbf{Dynamic Reference Utilization:} In a deployed RAG system, a query might be supported by a dynamic number of useful references rather than one uniquely correct document. This flexibility is essential for capturing the inherent ambiguity and complexity of natural language queries. Shown in Figure \ref{fig:reference_distribution}, while a fixed retrieval method might return a preset number of documents, our CAR method dynamically adjusts the number of retrieved references based on query complexity, thereby balancing efficiency with contextual richness.

\textbf{Experimental Setup and Methodology:}
In our experiments, we evaluate the performance of our proposed cluster-based adaptive retrieval algorithm, CAR, within a Retrieval-Augmented Generation (RAG) framework deployed on the CDP search bar. We compare CAR against two fixed retrieval baselines: Top‑40 and Top‑16. In the Top‑40 baseline, every query retrieves 40 documents (30 from CDP documentation and 10 from API references), whereas the Top‑16 baseline consistently retrieves 16 documents (12 from CDP documentation and 4 from API references). Top-40 threshold was chosen based on prior experiments, which demonstrated that retrieving at least 40 documents was necessary for the semantic search to consistently include all relevant results. The choice of 16 was guided by the CAR algorithm for reference, which retrieves an average of approximately 16 documents.

CAR is initialized with the same set of 40 candidate documents used in the Top‑40 baseline. It then applies a clustering-based selection process to filter out less relevant documents, retaining only those that best match the specific information needs of the query. As a result shown in table \ref{tab:retrieval_efficiency}, CAR returns, on average, approximately 16 documents per query, effectively tailoring the retrieved context to each query.

We conducted our experiments on 1400 real-world queries collected from the CDP search bar. The primary retrieval metrics evaluated include \textit{the number of retrieved documents}, \textit{the token count} and \textit{end-to-end pipeline latency}. To assess the quality of responses generated using different retrieval strategies, we employ the \textit{DeepEval} open-source library\footnote{GitHub URL: \url{https://github.com/confident-ai/deepeval}}, focusing on the following metrics: 
\begin{itemize} [leftmargin=9pt]
\item \textbf{Answer Relevancy:} Measures the alignment of the generated response with the query’s intent and correctness. 
\item \textbf{Hallucination:} Quantifies the presence of incorrect or fabricated information in the response by cross-checking the generated content against the retrieved context. A hallucination score of 1 indicates the presence of fabricated content, whereas a score of 0 indicates its absence. 
\item \textbf{Contextual Relevancy:} Evaluates the extent to which the retrieved context supports the given query, defined as: \begin{equation}\nonumber \text{Contextual Relevancy} = \frac{\text{Number of Relevant Statements}}{\text{Total Number of Statements}} \end{equation} 
\end{itemize}

\textbf{Experimental Findings and Discussion}:
Figure \ref{fig:reference_distribution} illustrates the distribution of reference counts retrieved by different retrieval methods, highlighting the dynamic nature of CAR based on the complexity of the query. CAR exhibits a more varied distribution compared to static methods, adapting the number of retrieved references dynamically based on the complexity of the query. This adaptability allows CAR to balance efficiency and relevance by fetching more references for complex queries while reducing unnecessary retrievals for simpler ones.

Table~\ref{tab:retrieval_efficiency} presents a performance comparison of the Top‑40, Top‑16, and CAR methods, averaged over 1400 queries. CAR always retrieves substantially fewer documents than the Top‑40 baseline, leading to a significant reduction in total token consumption by 60\%. This also translates to reduction in end-to-end LLM generation latency and costs.

In terms of response quality, CAR demonstrates strong performance in both answer relevancy and Contextual Relevancy while significantly reducing the hallucination rate by 10\% compare to Top-40 baseline. \textbf{As a financial institution, Coinbase must prioritize response accuracy, as even a slight increase in hallucinations can degrade user experience and pose reputational risks.} Although the Top‑40 method achieves marginally higher answer relevancy, this advantage is offset by its increased hallucination rate, likely due to retrieving excessive and potentially redundant information. Due to its high hallucination rate, Top-16 is the worst performing method from a deployed application perspective.

Overall, these findings validate CAR’s effectiveness in balancing retrieval efficiency with high-quality response generation. By optimizing document selection and minimizing hallucinations, CAR offers a simple, scalable and robust solution for real-world RAG deployments.

\begin{table}[htbp]
\centering
\resizebox{0.45\textwidth}{!}{
\begin{tabular}{|l|c|c|}
\hline
\textbf{Clustering Method} & \textbf{Coinbase CDP} & \textbf{MultiHop-RAG} \\
\hline
K-Means \cite{macqueen1967some} & 0.62 & 0.33 \\
DBSCAN \cite{ester1996density} & 0.62 & 0.31 \\
HDBSCAN \cite{mcinnes2017hdbscan} & 0.62 & 0.30 \\
OPTICS \cite{ankerst1999optics}& 0.58 & 0.32 \\
Agglomerative \cite{ward1963hierarchical} & 0.62 & 0.33 \\
Spectral \cite{ng2001spectral}& 0.58 & 0.32 \\
BIRCH \cite{zhang1996birch} & \textbf{0.63} & 0.33 \\
Bisecting KMeans \cite{karypis2000comparison}& 0.59 & \textbf{0.34} \\
\hline
\end{tabular}
}
\caption{Comparison of CAR performance measured by  Trade-off Efficiency Score (TES; higher is better) with different clustering algorithm on Coinbase CDP and MultiHop-RAG Datasets}
\label{tab:car_comparison}
\end{table}

\vspace{-3mm}
\subsection{Effect of Clustering Algorithm Choice}

Table~\ref{tab:car_comparison} reports CAR’s Trade-off Efficiency Score when its adaptive cut-off is driven by eight popular clustering algorithms.  Results on the \textbf{Coinbase CDP} corpus are tightly clustered: K-Means, DBSCAN, HDBSCAN, and Agglomerative all reach 0.62, while OPTICS and Spectral trail slightly at 0.58.  The highest score, 0.63, is obtained with the BIRCH algorithm.

On the more heterogeneous \textbf{MultiHop-RAG} dataset, performance remains stable across methods, ranging from 0.30 to 0.34.  Bisecting K-Means delivers the best score (0.34), with K-Means, Agglomerative, and BIRCH close behind at 0.33.  Density-based approaches (DBSCAN and HDBSCAN) reach 0.31–0.32, showing only marginal degradation.

Overall, variation across clustering backbones is modest on both benchmarks, indicating that CAR’s gains stem primarily from the use of an adaptive cut-off rather than reliance on a specific clustering heuristic.
 
\vspace{-3mm}
\subsection{Clustering Parameter Optimization}
\label{hyperparameter-search}
For every clustering backbone we conduct a compact grid-search that maximises the
silhouette score of the scaled \textit{(index, distance)} pairs produced by CAR.
The search spaces are listed verbatim below for reproducibility, where \( n \) is the number of data points.

\textbf{K-Means} \cite{macqueen1967some}: The number of clusters (\textit{n\_clusters}) was optimized by searching integers from 2 up to a maximum determined by half the data size, capped by \( n \).

\textbf{DBSCAN} \cite{ester1996density}: The neighborhood distance (\textit{eps}) was tested across five steps from 0.1 to 1.0, while the minimum neighborhood size (\textit{min\_samples}) was tested for integers from 2 up to 5 (or \( n - 1 \) if smaller).

\textbf{HDBSCAN} \cite{mcinnes2017hdbscan}: The minimum cluster size (\textit{min\_cluster\_size}) was tested with values \{2, 3, 4, 5\} (up to \( n \)), and \textit{min\_samples} was tested with \{1, 2, 3\} (constrained to be less than \( n \) and no larger than \textit{min\_cluster\_size}).

\textbf{OPTICS} \cite{ankerst1999optics}: The minimum neighborhood size (\textit{min\_samples}) was tested with values \{2, 3, 4, 5\} (less than \( n \)), and the cluster extraction steepness (\textit{xi}) was tested with \{0.01, 0.05, 0.1\}.

\textbf{Agglomerative Clustering} \cite{ward1963hierarchical}: The number of clusters (\textit{n\_clusters}) was optimized similarly to K-Means, and the linkage method (\textit{linkage}) was tested using `ward', `average', and `complete'.

\textbf{Spectral Clustering} \cite{ng2001spectral}: The number of clusters (\textit{n\_clusters}) was optimized similarly to K-Means, using a dynamically set number of neighbors based on data size.

\textbf{BIRCH}\cite{zhang1996birch}: The number of clusters (\textit{n\_clusters}) was optimized similarly to K-Means, while the CF-tree threshold (\textit{threshold}) was tested with values \{0.3, 0.5, 0.7\}.

\textbf{Bisecting K-Means} \cite{karypis2000comparison}: The number of clusters (\textit{n\_clusters}) was optimized by searching integers from 2 up to a maximum determined by half the data size, capped by \( n \).

\vspace{-3mm}
\subsection{Empirical Case Study}
Next, we demonstrate that the CAR approach achieves the lowest hallucination rate with the following examples. Consider the user query \textbf{\textit{"adding to quicken mac"}}. The Top-40 retrieval strategy produces the following response:

\begin{tcolorbox}[
    colback=gray!10,  
    colframe=gray!50, 
    arc=3pt,          
    left=3pt, right=3pt, top=3pt, bottom=3pt, 
    boxsep=1pt        
]
\textit{You are asking about adding Coinbase transactions to Quicken for Mac. The Coinbase Developer Platform (CDP) does not directly support importing transactions into Quicken for Mac. Quicken for Mac does not have a built-in integration with Coinbase. \\
To import your Coinbase transactions into Quicken for Mac, you can use a third-party tool like [Coinbase to Quicken](https://www.coinbasetoquicken.com/). This tool allows you to download your Coinbase transaction history and import it into Quicken for Mac.}
\end{tcolorbox}

This response contains incorrect information and even includes a fabricated link. In contrast, CAR retrieves more relevant context, enabling the LLM to generate a more accurate and reliable response while utilizing a more efficient input context:

\begin{tcolorbox}[
    colback=gray!10,  
    colframe=gray!50, 
    arc=3pt,          
    left=3pt, right=3pt, top=3pt, bottom=3pt, 
    boxsep=1pt        
]
\textit{You are asking about adding Coinbase Exchange data to Quicken for Mac. The Coinbase Developer Platform (CDP) does not currently offer direct integration with Quicken for Mac. Quicken is a personal finance software, and CDP focuses on providing APIs for developers to build applications and integrate with Coinbase services.\\
To manage your Coinbase Exchange transactions in Quicken for Mac, you can manually import your transaction history from the Coinbase Exchange website.}
\end{tcolorbox}

Let's consider another example to illustrate how top-40 results in hallucination due to inundated context. Consider the user query \textbf{\textit{"fetching recurring buys"}}. The response generated using the Top-40 retrieval strategy is:
\begin{tcolorbox}[
    colback=gray!10,  
    colframe=gray!50, 
    arc=3pt,          
    left=3pt, right=3pt, top=3pt, bottom=3pt, 
    boxsep=1pt        
]
\textit{"The CDP API does not currently support recurring buys. The Coinbase App v2 Buys API has been deprecated and no longer supports recurring buys."}
\end{tcolorbox}

This response is factually incorrect. Specifically, the claim \emph{``no longer supports recurring buys"} is misleading as the CDP API never supported recurring buys. This error arises due to excessive and irrelevant documents in the retrieval, preventing the LLM from focusing on the most important information--an issue known as the ``Lost in the Middle" phenomenon \cite{liu2024lost}.

In contrast, the dynamic CAR approach retrieves more relevant documents, resulting in better response quality while avoiding the pitfalls of excessive or irrelevant retrieval, as show below.
\begin{tcolorbox}[
    colback=gray!10,  
    colframe=gray!50, 
    arc=3pt,          
    left=3pt, right=3pt, top=3pt, bottom=3pt, 
    boxsep=1pt        
]
\textit{The Coinbase Developer Platform (CDP) API does not directly support recurring buys. The [Wallet API: Buys](https://docs.cdp.coinbase.com/coinbase-app/docs/api-buys) endpoint, which was deprecated on November 30, 2023, allowed for placing buy orders. However, it did not offer functionality for recurring buys.}
\end{tcolorbox}

The Top-16 retrieval strategy exhibits the highest hallucination rate while also yielding lower answer relevancy. This outcome is expected, as complex retrieval-augmented generation (RAG) pipelines—such as those supporting CDP documentation—require retrieving sufficient context. A lower top-\(k\) value often omits critical information, leading to incomplete or incorrect responses.

For example, consider the user query \textbf{\textit{"download historical spot price of btc-eur"}}. The response generated using the Top-16 retrieval strategy is:

\begin{tcolorbox}[
    colback=gray!10,  
    colframe=gray!50, 
    arc=3pt,          
    left=3pt, right=3pt, top=3pt, bottom=3pt, 
    boxsep=1pt        
]
\textit{The Coinbase Developer Platform (CDP) provides various APIs for accessing market data, including historical spot prices. However, the specific endpoint for downloading historical spot prices is not explicitly documented in the provided URLs.}
\end{tcolorbox}

This response lacks specificity and fails to direct the user to the relevant API. In contrast, CAR dynamically adjusts to a larger K value, retrieving the appropriate documentation and enabling the LLM to generate a more informative and precise response (shortened due to space constraints):

\begin{tcolorbox}[
    colback=gray!10,  
    colframe=gray!50, 
    arc=3pt,          
    left=3pt, right=3pt, top=3pt, bottom=3pt, 
    boxsep=1pt        
]
\textit{The Coinbase Developer Platform (CDP) API provides various endpoints for retrieving market data, including historical spot prices.}\\
1. \href{https://docs.cdp.coinbase.com/coinbase-app/docs/api-prices}{Get Spot Price}: \textit{This endpoint allows you to retrieve the current market price for a given currency pair.} 

2. \href{https://docs.cdp.coinbase.com/exchange/docs/fix-msg-market-data#marketdatasnapshotfullrefresh-35w}{Exchange FIX Market Data Messages}: \textit{The Coinbase Exchange FIX API provides a comprehensive set of market data messages, including historical spot prices. }

\end{tcolorbox}


Overall, our experiments demonstrate that CAR effectively optimizes retrieval by dynamically adjusting the number of retrieved documents. It reduces search costs and latency while maintaining high answer quality, making it a practical alternative to static top-\(k\) retrieval methods.

\section{Conclusion}
\label{sec:conclusion}
We introduced Cluster-based Adaptive Retrieval (CAR), a novel method for dynamically selecting documents in RAG pipelines. CAR improves upon fixed top-\(k\) strategies by leveraging clustering techniques to determine an optimal cut-off, ensuring relevant context while reducing redundancy and computational overhead. 
By optimizing document selection, CAR sets a new standard for intelligent and efficient RAG pipelines.
Our experiments on the CDP documentation demonstrate the superiority of CAR over traditional methods by reducing token consumption, latency, and hallucination. Case studies from production deployments highlight its adaptability across query complexities, effectively balancing completeness and efficiency.

\bibliographystyle{ACM-Reference-Format}
\bibliography{bibli}

\clearpage

\appendix

\end{document}